\documentclass[aip,jcp,reprint,amsmath,amssymb,floatfix,citeautoscript]{revtex4-1}
\usepackage{amsmath}
\usepackage{graphicx}
\usepackage{amssymb}
\usepackage{amsthm}
\usepackage{bm}

\usepackage{ragged2e}
\usepackage[labelsep=period,format=plain,justification=justified,font=footnotesize]{caption}
\DeclareCaptionJustification{myjust}{\justifying}
\usepackage{times}
\usepackage{txfonts}

\newcommand{\vr}{{\mathbf{r}}}

\begin{document}

\title{Microscopic theory of singlet exciton fission. II. Application to pentacene dimers and the role of superexchange
}

\author{Timothy C. Berkelbach}
\email{tcb2112@columbia.edu}
\affiliation{Department of Chemistry, Columbia University, 3000 Broadway, New York, New York 10027, USA}

\author{Mark S. Hybertsen}
\email{mhyberts@bnl.gov}
\affiliation{Center for Functional Nanomaterials, Brookhaven National Laboratory, Upton, New York 11973-5000, USA}

\author{David R. Reichman}
\email{drr2103@columbia.edu}
\affiliation{Department of Chemistry, Columbia University, 3000 Broadway, New York, New York 10027, USA}

\begin{abstract}

In the preceding paper, we assembled the theoretical components necessary for a unified framework of singlet fission, a
type of multiexciton generation producing two triplet excitons from one singlet exciton.  In this paper, we apply our methodology
to molecular dimers of pentacene, a widely studied material that exhibits singlet fission.  We address a longstanding
theoretical issue, namely whether singlet fission proceeds via two sequential electron transfer steps mediated by a charge-transfer
state or via a direct two-electron transfer process.  We find evidence for a superexchange mediated mechanism, whereby
the fission process proceeds through virtual charge-transfer states which may be very high in energy.  In particular, this mechanism
predicts efficient singlet fission on the sub-picosecond timescale, in reasonable agreement with experiment.  We investigate 
the role played by molecular vibrations in mediating relaxation and decoherence, finding that different physically reasonable forms
for the bath relaxation function give similar results.  We also examine the competing direct coupling mechanism and find it to yield 
fission rates slower in comparison with the superexchange mechanism for the dimer. We discuss implications for crystalline pentacene,
including the limitations of the dimer model.

\end{abstract}

\maketitle

\section{Introduction}

Singlet exciton fission, a process whereby one high-energy singlet exciton is converted into two lower energy triplet excitons,
is an excited state phenomenon with a potential impact on the efficiency of inexpensive organic solar cells\cite{smi10}.
In light of this potential utility in the design of photovoltaic systems,
there has been a recent explosion of experimental studies on a variety of molecular
materials\cite{joh10,bur10,bur11,cha11,wil11,rob12,cha12,ma12,ram12} and fabricated devices\cite{lee09,jad11,ehr12}. 
Currently, a detailed microscopic understanding of this process is lacking.
In the preceding paper\cite{ber12_sf1}, we outlined a fully microscopic theoretical framework for the practical
simulation of singlet fission dynamics.  In particular, within the context of singlet fission chromophore systems,
we connected excited state quantum chemistry with established
reduced density matrix methods from quantum relaxation theory.

Here, we continue this endeavor with a realistic theoretical treatment of the singlet fission dynamics of molecular dimers.
Understanding singlet fission in small molecular complexes has intrinsic interest and such complexes may have utility 
as a sensitizer in a Gr{\" a}tzel-type solar cell\cite{gra03,pac06}. They may also be representative model systems to understand the singlet 
fission process in bulk crystals.  We focus on pentacene because it is perhaps the most thoroughly studied material that 
has been robustly shown to exhibit singlet fission. In particular, experimental evidence points to fast and efficient 
singlet fission in bulk pentacene\cite{mar07,mar09,joh09,cha11,wil11},
but the process appears to be much slower in the molecular (tetracene) complexes realized so far\cite{mul06,mul07}.
Here we will explore dimer systems both as a realistic description of small molecular complexes and as
a model potentially applicable in bulk materials.
Full application of the present formalism to pentacene clusters and crystals will be subject of a future report.

\begin{figure}[b]
\centering
\includegraphics[scale=1.0]{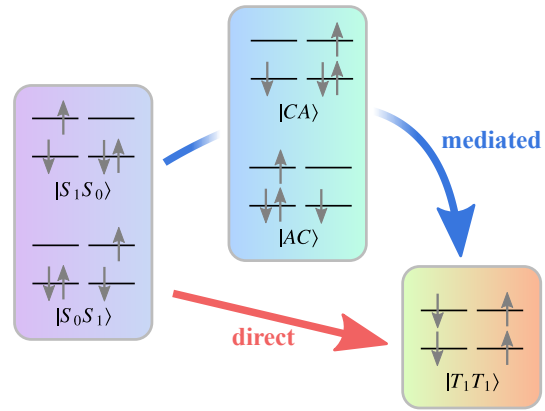}
\caption{Schematic representation of the five electronic states relevant for singlet fission in a dimer.  The actual
states employed in the calculations are spin-adapted linear combinations yielding overall spin-singlets, unlike
those shown here.}
\label{fig:states}
\end{figure}

An ongoing debate, the crux of which is laid out in Ref.~\onlinecite{smi10} and in our previous paper\cite{ber12_sf1}, concerns the
quantum mechanical mechanism by which singlet fission takes place amongst the manifold of excited states.
Specifically, this debate rests on the role, or lack thereof,
of charge-transfer (CT) states in the singlet fission process.  The relevant quantum mechanical states as well as the two competing
mechanisms are depicted schematically in Fig.~\ref{fig:states}. The so-called ``mediated'' mechanism, posits that a single electron
transfer from the intramolecular excited singlet state, $|S_1S_0\rangle$ or $|S_0S_1\rangle$, produces a charge-transfer state, $|CA\rangle$ or $|AC\rangle$,
after which a second one-electron transfer event yields the multi-exciton, triplet-triplet state, $|T_1T_1\rangle$.  Alternatively,
a ``direct'' mechanism implies a simultaneous two-electron process which circumvents CT states altogether.

Previous theoretical work on singlet fission has largely considered only the mediated mechanism.  In particular, Greyson et al.
investigated the case of purely coherent energy transfer for mediated fission\cite{gre10_jpcb1} and Teichen and Eaves
derived solvent-dependent rate expressions for the separate one-electron transfer events implicit in the mediated mechanism\cite{tei12}.
The latter authors concluded that a necessary condition for efficient mediated fission is that the CT state energy lie in between
that of the intramolecular singlet and the multiexciton triplet-triplet state, i.e. $E(S_1S_0), E(S_0S_1) > E(CA), E(AC) > E(T_1T_1)$.

This viewpoint is consistent with a recent quantum chemistry calculation on pentacene clusters reported by Zimmerman et al.\cite{zim11}. They concluded
that because CT states were calculated to be significantly higher in energy than intramolecular singlets (by about 300 meV or more), singlet fission
in pentacene cannot take place via the mediated mechanism.  Rather, the authors supported the direct mechanism, estimating a
direct coupling matrix element of about 5 meV.  However, this number is almost two orders of magnitude smaller than that required to explain
the experimentally observed timescale of fission in pentacene, ranging from 80 -- 200 fs\cite{mar07,mar09,joh09,cha11,wil11}.

The present manuscript is a first step towards the resolution of this apparent paradox.  Because singlet fission is inherently a dynamical
process, one must exercise caution in the interpretation of \textit{static} electronic structure calculations and their implications for
fission.  As such, we argue that a microscopic,
dynamical treatment of the relevant electronic states coupled to a finite temperature bath is crucial for a
theoretically sound description of singlet fission processes.  Furthermore, the accuracy of the methodology and its associated approximations must
be established for these complex problems.  The implementation should be carefully benchmarked and thoughtfully parametrized
for the relevant physical problem, in this case, singlet fission.  We have carried out the first step of benchmark calculations in our previous
paper and here we take the second step,
parametrizing a system-bath Hamiltonian for fission in molecular dimers and using an accurate quantum relaxation
master equation to calculate the fission dynamics.  Through this program, we are able to make firm statements regarding the feasibility
of competing mechanisms as well as predict and rationalize experimental fission rates.

The layout of the paper is as follows.  We begin in Sec.~\ref{sec:method} with a review of the methodology presented in our previous
paper\cite{ber12_sf1}.  In Sec.~\ref{sec:results} we present our results for pentacene, which explore the effects of
energy levels, electronic couplings, and phonon properties.  Although we use pentacene as an example molecule, our exploration of important
singlet fission parameters is sufficiently broad so as to elucidate generic aspects of singlet fission.  We summarize our work
and conclude in Sec.~\ref{sec:conc}.

\section{Methodology}\label{sec:method}

In this section, we briefly describe the adopted theoretical methodology as laid out in our previous paper, to which the reader
is referred for more details\cite{ber12_sf1}.  
In essence, we employ a system-bath Hamiltonian describing the coupling of the electron and phonon degrees of freedom\cite{bre02,may11},
$\hat{H}_{tot} = \hat{H}_{el} + \hat{H}_{el-ph} + \hat{H}_{ph}$, with
\begin{equation}
\hat{H}_{el} = \sum_i |i\rangle E_i \langle i| + \sum_{i\neq j} |i\rangle V_{ij} \langle j |,
\end{equation}
\begin{equation}
\hat{H}_{el-ph} = \sum_i |i\rangle\langle i | \sum_{k} c_{k,i} \hat{q}_k
    + \sum_{i\neq j} |i\rangle \langle j | \sum_{k} c_{k,ij} \hat{q}_k,
\end{equation}
and
\begin{equation}
\hat{H}_{ph} = \sum_{k} \left[ \frac{\hat{p}_k^2}{2} + \frac{1}{2}\omega_k^2 \hat{q}_k^2 \right].
\end{equation}
The parameters of this Hamiltonian are determined via a variety of \textit{ab initio} and semi-empirical methods, and the
dynamics generated under the action of this Hamiltonian are calculated by a perturbative quantum master equation.
The electronic structure and quantum dynamics methodologies are described in the following sections.

\subsection{Geometry and electronic structure}

\begin{figure}[b]
\centering
\includegraphics[scale=1.0]{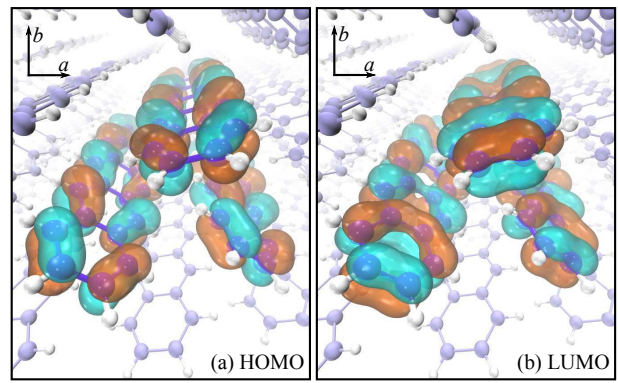}
\caption{
Molecular geometry of the pentacene crystal.  Three pentacene molecules are emphasized, displaying the three
symmetry-unique nearest-neighbor dimer pairs discussed in the text.
Also shown are isosurface plots of the HF HOMO (a) and LUMO (b) of the isolated molecules, including the phase
convention adopted in this work.  
}
\label{fig:dimers}
\end{figure}

To evaluate the role of molecular geometry, we consider individual pentacene dimers extracted from the $ab$ plane of the
experimental crystal structure\cite{mat01}.  
There are approximately three symmetry unique nearest-neighbor dimer pairs, as shown in Fig.~\ref{fig:dimers}, corresponding to the $\left[a\ b\right]$
translation
vectors $\left[1\ 0\right]$, $\left[1/2\ 1/2\right]$, and $\left[-1/2\ 1/2\right]$.  By investigating these dimers in particular, we are probing
the extent to which efficient singlet fission observed in crystals but not dimers may be due simply to molecular orientation as opposed to explicit
many-body environmental effects.
Alternatively, for a different choice of electronic structure parameters (which may be altered by electronic
polarization for example), the dimers may be taken as model
systems for the crystal, assuming purely local energy transfer events as described in the introduction.
Though such a picture permeates the literature, one should not invest too much
in this model-based view, instead preferring a direct treatment of the full system as done in our forthcoming work on pentacene crystals. Exciton
delocalization in neat acene crystals may span up to tens of molecules\cite{cha12}, invalidating this simple dimer picture.
Before concluding, we will discuss analogous calculations on a covalently linked pentacene dimer inspired by previously
studied tetracene dimers\cite{mul06,mul07} with the aim of rationalizing the low observed fission yield.

When applied to a molecular dimer, the electronic model Hamiltonian described in our previous paper yields five diabatic states (excluding the
ground state), each one a spin singlet.
The first two are localized,
intramolecular Frenkel excitations, $|S_1 S_0\rangle$ and $|S_0 S_1\rangle$.  Additional single excitations generate the third and fourth states
of charge-transfer character, $|CA\rangle$ and $|AC\rangle$.  The fifth and final state is a doubly excited triplet-triplet state $|T_1T_1\rangle$,
presumed to be the spin singlet precursor to fully separated triplets.  These five states are depicted schematically in Fig.~\ref{fig:states}, though
we emphasize that the single-configuration states shown there are not states of well-defined spin multiplicity, i.e. they are not eigenfunctions
of the $S^2$ operator.
The spin-adapted variant of these dimer states employed here (as well as their non-spin-adapted counterparts) have been used as a starting point in a
variety of other theoretical works\cite{smi10,gre10_jpcb1,tei12}.

Because calculated excited state energies exhibit errors on the order of 0.5 eV, especially for acenes\cite{kad06},
we take the diagonal matrix elements of the electronic Hamiltonian,
$E_i$, to be adjustable parameters.  This procedure avoids any bias inherited from electronic structure methodology and more importantly provides for
qualitative, physical insight into the effect that electronic energies have on singlet fission dynamics.  The electronic couplings, $V_{ij}$, on the other
hand, are taken from ab initio calculations using the Hartree-Fock (HF) molecular orbitals (MOs) of isolated pentacene molecules.  As laid out
in App. A of our previous paper and Eqs.~(7)-(13) of Ref.~\onlinecite{smi10}, one-electron couplings are given by off-diagonal elements
of the Fock operator for the combined, two-molecule system.  Specifically, we have
\begin{align}
\langle CA | \hat{H}_{el} | S_1 S_0 \rangle &= t_{LL}, & \langle AC | \hat{H}_{el} | S_1 S_0 \rangle &= -t_{HH}, \nonumber\\
\langle CA | \hat{H}_{el} | T_1 T_1 \rangle &= \sqrt{3/2}\ t_{LH}, & \langle AC | \hat{H}_{el} | T_1 T_1 \rangle &= \sqrt{3/2}\ t_{HL}, \nonumber\\
\langle CA | \hat{H}_{el} | S_0 S_1 \rangle &= -t_{HH}, & \langle AC | \hat{H}_{el} | S_0 S_1\rangle &= t_{LL}. \nonumber
\end{align}
In the above, $t_{LL}$ and $t_{HH}$ denote the one-electron coupling
of the LUMO and HOMO, respectively, whereas $t_{LH}$ is the
electronic coupling between the LUMO of the first molecule (A) and the HOMO of the second (B) and likewise for $t_{HL}$, i.e.
\begin{align}
t_{LL} &= \langle L_A | \hat{F} | L_B \rangle, & t_{HH} &= \langle H_A | \hat{F} | H_B \rangle, \nonumber\\
t_{LH} &= \langle L_A | \hat{F} | H_B \rangle, & t_{HL} &= \langle H_A | \hat{F} | L_B \rangle. \nonumber
\end{align}
It should be noted that the mixed
couplings, $t_{LH}$ and $t_{HL}$, are entirely responsible for CT-mediated singlet fission.
Thus, whereas previous theoretical studies of singlet fission
have invoked Longuet-Higgins-type approximations to estimate the mixed couplings\cite{gre10_jpcb1},
the direct evaluation in terms of the Fock operator employed here should
be preferred.

Additionally, the direct evaluation yields the {\em sign} of all electronic couplings, in contrast to
approximate methods (such as the energy-splitting in a dimer method\cite{gre10_jpcb2}) that yield only the magnitude of the
coupling (however the sign can often be inferred by inspection of the orbitals).  In multistate
systems and band theory calculations, the sign of the coupling can be very important and should be retained whenever possible.
To achieve a consistent sign, one must adopt a phase convention for the molecular orbitals of the system.  In agreement with previous
studies\cite{yam11}, we employ an approximate screw-axis to fix the phase of the MOs, which are shown as used in Fig.~\ref{fig:dimers}.

In addition to the one-electron coupling matrix elements, a variety of two-electron integrals are apparent in the coupling expressions
given in our previous paper and Ref.~\onlinecite{smi10}.  These two-electron integrals, when included, were calculated by representing the HF MOs
on a real-space grid and subsequently utilizing fast Fourier transform techniques\cite{bae10}.  Integrals were checked for convergence with respect
to the grid size.

All quantum
chemistry calculations were performed with either the GAMESS (US) quantum chemistry package\cite{gamess} or the 
Firefly quantum chemistry package\cite{firefly}, which is partially based on the GAMESS (US) source code. Calculations
employed the 6-31G(d) basis set.

\subsection{Quantum dynamics}

Reduced density matrix (RDM) quantum dynamics calculations were performed within the Redfield framework\cite{blu81,bre02,may11,ish09_jcp1,pol96},
with the secular and Markov approximations, whose use was justified
theoretically and numerically in our previous paper\cite{ber12_sf1} to which we refer the reader for technical details.  Briefly, the electronic
RDM, which follows from a trace of the total density matrix, $W(t)$, over the phonon degrees of freedom,
$\rho_{ij}(t) = \langle i | \textrm{Tr}_{ph} W(t) |j \rangle$, describes the time evolution of the populations, $\rho_{ii}\equiv P_i$,
and coherences, $\rho_{ij}$, of the electronic states.  In the adiabatic basis that diagonalizes the electronic Hamiltonian,
$\hat{H}_{el} |\alpha\rangle = \hbar \omega_\alpha |\alpha\rangle$, the RDM obeys the equation of motion
\begin{equation}\label{eq:redfield}
\frac{d\rho_{\alpha\beta}(t)}{dt} = -i\omega_{\alpha\beta}\rho_{\alpha\beta}(t)
    + \sum_{\gamma,\delta} R_{\alpha\beta\gamma\delta} \rho_{\gamma\delta}(t).
\end{equation}
The Redfield tensor\cite{bre02,ish09_jcp1}, $R_{\alpha\beta\gamma\delta}$,
which effects finite-temperature relaxation and dephasing processes, can be expressed
in terms of thermal bath correlation functions.
Those correlation functions are entirely determined by the spectral densities of the phonon
degrees of freedom,
\begin{equation}
J_{ii}(\omega) =  \frac{\pi}{2} \sum_{k} \frac{c_{k,i}^2}{\omega_k} \delta(\omega-\omega_k)
\end{equation}
and
\begin{equation}
J_{ij}(\omega) = \frac{\pi}{2} \sum_{k} \frac{c_{k,ij}^2}{\omega_k} \delta(\omega-\omega_k),
\end{equation}
which may be calculated through a combination of molecular dynamics and quantum chemistry calculations.
Thus, the form of the spectral density adopted in a RDM calculation encapsulates the complicated detailed
motion of the vibrational phonon degrees of freedom.  For simplicity, most of our results will employ
the common Ohmic spectral density with a Lorentzian high-frequency cutoff,
\begin{equation}\label{eq:ohmic}
J^{O}_{ij}(\omega) = 2 \lambda_{ij} \Omega_{ij} \omega \frac{1}{\omega^2 + \Omega_{ij}^2},
\end{equation}
with the the strength of the system-bath interaction quantified by the reorganization energies
$\lambda_{ij} = \pi^{-1} \int d\omega J_{ij}^{O}(\omega)/\omega$, and the
frequency of the interaction quantified by the cutoff $\Omega_{ij}$.  
To investigate the crucial role played by phonons
in our dynamical calculations, we will also consider two alternative forms of the spectral density,
to be discussed in Sec.~\ref{ssec:bath} and shown in Fig.~\ref{fig:spec}.  This latter study highlights a particularly important advantage
of the Redfield formalism, because it can easily be applied to any form of the spectral density, unlike
many other methods.

\section{Results for pentacene}\label{sec:results}

\begin{table}[t]
\begin{tabular*}{0.48\textwidth}{@{\extracolsep{\fill}} lcccc }
\hline\hline
$[a\ \ b]$            & $t_{HH}$      & $t_{LL}$      & $t_{HL}$      & $t_{LH}$\\
\hline
$[1\ \ 0]$       & 85 (34,51)    & -60 (-43)     & -74           & 74     \\
$[1/2\ \ 1/2]$   & -145 (47,-74) & 116 (-82)     & 109           & -124    \\
$[-1/2\ \ 1/2]$  & 228 (-85,131) & -111 (84)     & 108           & -134    \\
\hline\hline
\end{tabular*}
\caption{Electronic coupling parameters (in meV) of pentacene for the three dimer types 
described in the text.  Values in parentheses are those calculated by Yamagata et al.\cite{yam11} and Troisi and Orlandi\cite{tro05},
the latter only where available ($t_{HH}$).}
\label{tab:coupling}
\end{table}

We now proceed to apply the above methodology to pentacene.  The one-electron coupling parameters
calculated as described are given in Tab.~\ref{tab:coupling}, and are in reasonable agreement with values
obtained by semiempirical calculations\cite{yam11} and density functional theory (DFT)\cite{tro05}.  All electronic
coupling values are clearly on the 100 meV order of magnitude, though we point out that HF appears to systematically
yield larger couplings as compared to other methods.  A uniform rescaling may be performed as in Sec.~\ref{ssec:sxscaling},
but we will not do so here.  Henceforth, we will only present results for the
[1/2 1/2] dimer, but results for the other two are qualitatively similar, with quantitatively different dynamical timescales.
As discussed above, the diagonal energies will be varied in our simulations, the results of which are presented in Sec.~\ref{ssec:energy}.
Furthermore, we initially neglect all two-electron integrals, thereby investigating only the mediated fission mechanism.
Combining the computed one-electron couplings with the aforementioned expressions thus yields the electronic Hamiltonian
in units of meV,
\begin{equation}
\hat{H}_{el} =
\left(
\begin{array}{ccccc}
E(CA)   & 116       & -152      & 145       & 0     \\
116     & E(S_1S_0) & 0         & 0         & 145   \\
-152    & 0         & E(T_1T_1) & 0         & 133   \\
145     & 0         & 0         & E(S_0S_1) & 116   \\
0       & 145       & 133       & 116       & E(AC)
\end{array}
\right).
\end{equation}

The phonon bath will be characterized by only diagonal system-bath coupling, initially of the Ohmic form, Eq.~(\ref{eq:ohmic}),
with $J_{i\neq j}(\omega) = 0$, i.e. we exclude the off-diagonal Peierls coupling, but will revisit this topic in Sec.~\ref{ssec:bath}.
We have performed the calculations presented here
in the presence of off-diagonal coupling to a low-frequency bath and find the results to be largely unchanged; such low-frequency
modes are simply not efficient at mediating large-scale energy transfer, i.e. $\hbar \Omega_{ij} \ll |E_i-E_j|$.  For simplicity, we assume identical, uncorrelated baths
for each electronic state, i.e. $\Omega_{ii} \equiv \Omega$ and $\lambda_{ii} \equiv \lambda$.  In reality, the bath parameters will
be slightly different for each state and perhaps correlated because they share certain molecules.  While we surmise that these
effects will only be of quantitative significance, they are a potentially interesting topic for further research.  The cutoff frequency is
$\hbar \Omega = 180$ meV $= 1450$ cm$^{-1}$, as is typical for acenes and other conjugated organic molecules.
The reorganization energy is $\lambda = 50$ meV $= 400$ cm$^{-1}$. Though these parameters imply a Huang-Rhys factor,
$S = \lambda/\hbar \Omega \approx 0.3$, which is about a factor of two lower than the experimental one (see our previous paper for a theoretical
comparison to the experimental vibronic spectrum of Ref.~\onlinecite{tao07}), they insure a quantitatively
accurate treatment of dynamics within the weak-coupling, Redfield framework, as demonstrated numerically in our previous paper.  In
Sec.~\ref{ssec:bath}, we will consider larger, more realistic values of the reorganization energy, for which the Redfield theory predictions
are still expected to be qualitatively accurate.

The initial condition of the reduced density matrix is $\rho(0) = |S_1 S_0\rangle\langle S_1 S_0|$, i.e. only one
molecule is initially excited.  Alternative initial conditions, such as a coherent or incoherent 
mixture of $|S_1 S_0\rangle$ and $|S_0 S_1\rangle$, were found to yield qualitatively similar results.  In most simulations,
we find that this $S_1$ superposition is achieved, regardless of the initial condition, on a 10 fs timescale, which is
much faster than the overall fission process.

\subsection{Scanning energies}\label{ssec:energy}

\begin{figure}[t]
\centering
\includegraphics[scale=0.33]{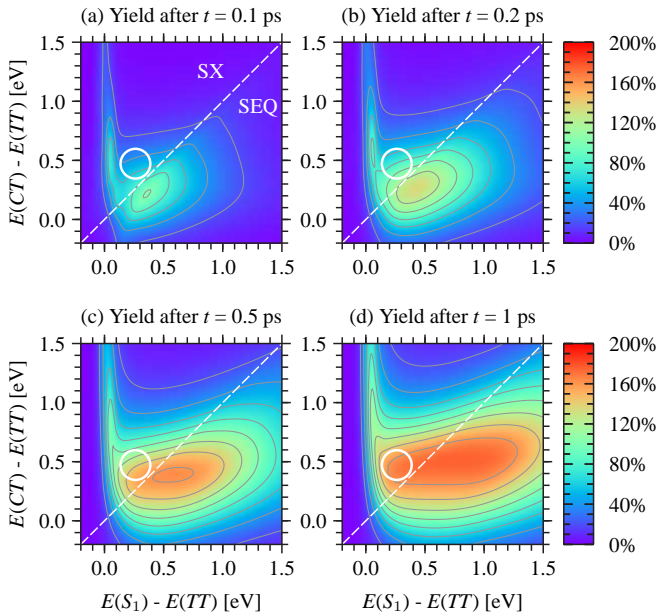}
\caption{
Singlet fission yield, $P_{TT}(t) \times 200\%$, after the four periods of time indicated for
the $\left[1/2\ 1/2\right]$ pentacene dimer.  The dashed line qualitatively separates the superexchange (SX)
regime, $E_{CT} > E_{S_1}$, from the sequential (SEQ) regime, $E_{S_1} > E_{CT}$.  Estimated energy levels for the pentacene
dimer are denoted by the white circle.
}
\label{fig:2dyield}
\end{figure}

Because we only consider a homodimer and exclude the ground state, we can simplify our notation, collectively referring to
the $S_1S_0$ and $S_0S_1$ states as ``$S_1$'', $CA$ and $AC$ as ``$CT$'', and $T_1T_1$ as ``$TT$''.
In this case, there are only two independent energetic parameters,
which we take to be the energy offset of the $CT$ states with respect to the $TT$ state, $E(CT)-E(TT)$, and the analogous
offset of the $S_1$ states to the $TT$ state, $E(S_1)-E(TT)$.  We naturally expect that fission will take place as
long as both of these parameters are positive, such that $TT$ is the lowest energy state (an assumption well-founded for
pentacene, based on experiment and calculations).  This expectation is clearly
validated in Fig.~\ref{fig:2dyield}, which shows the singlet fission yield after the four periods of time,
$t = 0.1$, 0.2, 0.5, and 1 ps.  The fission yield at time $t$ is calculated simply as the population of the diabatic
$TT$ state, $P_{TT}(t)$, times 200\%, the latter factor indicating conversion of one exciton into two.
Note that this metric is a combination of the rate of fission as well as the thermodynamic equilibrium of the system.
One can easily imagine situations where the rate of fission may be fast but thermodynamic equilibrium does not
overwhelmingly favor the $TT$ state.  Different physical situations and technological applications will dictate
whether it is more desirable to extract fewer carriers due to fission at short times (wherein one would want to optimize
the rate only) or wait longer to extract more carriers (optimize the equilibrium).  Of course the combination,
i.e. rapid fission with high thermodynamic efficiency, is most ideal and may also be possible in some situations.

Returning to Fig.~\ref{fig:2dyield}, we divide the energetic phase space into two regions, $E(CT) > E(S_1)$
and $E(CT) < E(S_1)$, demarcated by a dashed white line.  Only the latter yields the rather obvious energetic pathway for mediated fission, i.e.
population flows from $S_1$ to $CT$, and then from $CT$ to $TT$.  We call this the ``sequential'' mediated mechanism.
The sequential mechanism can be clearly seen in Fig.~\ref{fig:2dyield} to yield very efficient singlet fission, even at short time.
Perhaps the ideal energetic configuration is achieved for $E(S_1) - E(TT) = 400$ meV and $E(CT) - E(TT) = 200$ meV, which yields 
about 150\% singlet fission after only 200 fs.

We now consider the opposite energetic regime, $E(CT) > E(S_1)$.  Although this regime naively suggests a barrier to singlet fission
(recall that the direct coupling term has been set to zero),
we see a remarkably high fission yield, even at short times, as long as $0 < E(S_1)-E(TT) \lessapprox 500$ meV.  With this criterion
satisfied, efficient singlet fission occurs even for $CT$ energies up to 1 eV above $TT$. 
We refer to this somewhat surprising result as the ``superexchange'' mediated mechanism, a phenomenon familiar from electron transfer
in magnetic and photosynthetic systems\cite{mak96,sim97,bix97}, and introduced in our preceding paper\cite{ber12_sf1} in the context of singlet fission.
Clearly near the boundary $E(CT)=E(S_1)$, the distinction between ``sequential'' and ``superexchange'' is not so sharp. However we
will continue to adopt these names, so as to imply that the dynamics are mostly characteristic of either one or the other, i.e. these
limiting forms provide a useful language for the discussion of competing effects in CT-mediated singlet fission.

One may naturally question the relevance of the above analysis to pentacene, asking what are the relevant energetic
parameters for a pentacene dimer?
In all panels of Fig.~\ref{fig:2dyield}, we have placed a circle that encompasses the estimated energy levels for pentacene (discussed
below), clearly placing
it in the superexchange-dominated regime.
Recent calculations on pentacene dimers by Greyson et al.\cite{gre10_jpcb1} using a combination of time-dependent DFT and constrained
DFT, found $E(S_1)-E(TT) = 240$ meV and $E(CT)-E(TT) = 354$ meV.  As pointed out, accurate electronic structure calculations of excited states,
including those with multiple excitations, can be difficult and so we also consider estimates based on experimental
measurements.  To a first approximation (which was also adopted by Greyson et al.), the energy of the multi-exciton state is simply twice the energy of the
lowest triplet state, $E(TT) \approx 2E(T_1) = 2\times 0.86$ eV $= 1.72$ eV, where 0.86 eV is the experimental $T_1$ excitation energy\cite{bur77}. 
The first singlet excitation energy of
a pentacene monomer is approximately $E(S_1) =$ 2.1--2.3 eV\cite{bie80,hei98} giving an energy offset of $E(S_1) - E(TT) \approx 400$ meV.
Diabatic charge transfer energies are difficult to determine experimentally, but estimates from (adiabatic) spectroscopic
measurements on crystals suggest values upwards of 2.3--2.5 eV\cite{seb81,yam11}, thereby predicting $E(CT)-E(TT) \approx 600$ meV or more.

Most importantly, there is little debate that charge transfer energies are always higher in energy than those of the
first excited singlet, such that pentacene lies unambiguously in the superexchange regime of CT-mediated singlet fission.
While we of course cannot definitively conclude that fission in pentacene occurs exclusively via CT-mediated superexchange,
(given the many approximations in our work and because we have not yet addressed the possibility of direct fission) we can say with certainty that
\textit{high-lying CT intermediate states do not preclude efficient CT-mediated fission}.

\begin{figure}[t]
\centering
\includegraphics[scale=1.0]{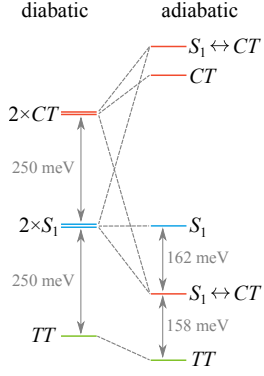}
\caption{
Energy level diagram depicting the diabatic electronic states (i.e. before mixing) and the adiabatic electronic states (i.e. after mixing),
for a typical ``superexchange'' energy configuration indicative of a pentacene dimer.
For adiabatic states which are a significant mixture of two different types of diabatic states, the notation $ i \leftrightarrow j$ is employed.
}
\label{fig:diabatic}
\end{figure}

\begin{figure}[t]
\centering
\includegraphics[scale=1.0]{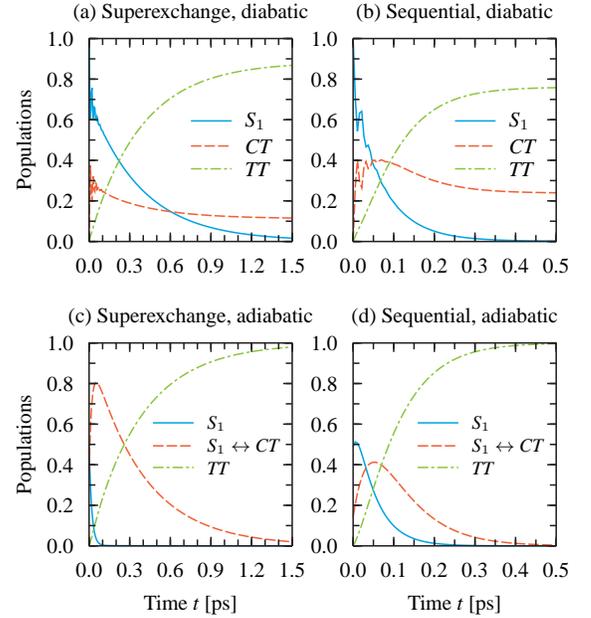}
\caption{
Population dynamics contrasting superexchange and sequential CT-mediated singlet fission, shown in both the
diabatic and adiabatic representations. Diabatic energy levels for panels (a) and (c) are $E(S_1)-E(TT)=250$ meV,
$E(CT)-E(TT)=500$ meV; and for panels (b) and (d) are reversed, i.e. $E(S_1)-E(TT)=500$ meV, $E(CT)-E(TT)=250$ meV.
}
\label{fig:dynamics}
\end{figure}

To visually summarize the results of this section, we first show in Fig.~\ref{fig:diabatic} an energy level diagram depicting the mixing of diabatic states to form
adiabatic states in the energetic arrangement $E(S_1)-E(TT) = 250$ meV and $E(CT)-E(TT) = 500$ meV which is approximately correct for pentacene and
characteristic of the superexchange regime.  The calculated population dynamics for this system are shown in Fig.~\ref{fig:dynamics} in both the diabatic (a) and
adiabatic (c) basis; the $S_1$ and $CT$ populations are given by $P_{S_1S_0}(t)+P_{S_0S_1}(t)$ and $P_{CA}(t)+P_{AC}(t)$ respectively.
In the diabatic basis one observes a very-short time mixing of $S_1$ and $CT$, after which $CT$ remains approximately constant while
$S_1$ decays into $TT$ with a single rate constant.  This behavior is exactly that of conventional superexchange, although the $CT$ population is slightly
larger than typical due to the strong electronic coupling.  The same behavior can be observed perhaps more directly in the adiabatic basis,
where the $S_1 \leftrightarrow CT$ superposition is population near-instantaneously, which then decays to an adiabatic state of essentially $TT$ character.
These dynamics should be contrasted with those of a sequential fission mechanism, with  $E(S_1)-E(TT) = 500$ meV and $E(CT)-E(TT) = 250$ meV,
shown in Fig.~\ref{fig:dynamics}(b) and (d).  In both bases it is clear
that a two-step kinetics prevails whereby an initially excited state first decays into an intermediate of $CT$ character,
which then itself decays into the final $TT$ state.  Though both mechanisms yield highly efficient singlet fission on the 1 ps timescale, their underlying
mechanistic details are clearly quite distinct.

\subsection{Superexchange and the strength of the electronic coupling}\label{ssec:sxscaling}

To understand how superexchange arises, consider the first-order effect that coupling to $CT$ states has upon the initially excited $S_1$ states,
\begin{equation}
|S_1^{(1)}\rangle \approx |S_1^{(0)}\rangle + \frac{V_{S_1,CT}}{E(S_1) - E(CT)} |CT^{(0)}\rangle
\end{equation}
from which the {\em effective} coupling from $S_1$ to $TT$ follows as
\begin{equation}\label{eq:supercoupling}
\begin{split}
\langle S_1^{(1)} | \hat{H}_{el} | TT^{(0)} \rangle &\approx \frac{V_{S_1,CT}V_{CT,TT}}{E(S_1)-E(CT)} \\
    &\hspace{-2em}= \frac{V_{S_1,CT}V_{CT,TT}}{\left[E(S_1)-E(TT)\right]-\left[E(CT)-E(TT)\right]}\\
    &\hspace{-2em}\approx \frac{-V_{S_1,CT}V_{CT,TT}}{\left[E(CT)-E(TT)\right]}.
\end{split}
\end{equation}
In the last line above we have assumed that $E(CT)-E(TT) \gg E(S_1)-E(TT)$.  For the electronic parameters of pentacene
considered above, this effective coupling is approximately 50 meV.
Performing second-order semiclassical (Marcus-like) perturbation theory in this effective electronic coupling yields
\begin{equation}
\begin{split}
k(S_1 \to TT) &\approx \frac{2\pi}{\sqrt{4\pi\hbar^2\lambda k_B T}}
    \frac{\left|V_{S_1,CT} V_{CT,TT} \right|^2}{\left[ E(CT)-E(TT)\right]^2} \\
    &\hspace{1em}\times \exp\left(-\frac{\left[ E(S_1)-E(TT)+\lambda \right]^2}{4\lambda k_B T}\right)
\end{split}
\end{equation}
from which one can directly read off a sharp, Gaussian dependence on the $S_1$ energy gap, with width
$\approx \sqrt{4\lambda k_B T} \approx 75$ meV, and a very weak, power-law dependence on the $CT$ energy gap,
both of which are consistent with the data previously shown in Fig.~\ref{fig:2dyield}.
Furthermore, it is apparent that the superexchange mechanism is a process which is overall fourth-order in the 
electronic coupling matrix elements, $V_{ij}$.

\begin{figure}[t]
\centering
\includegraphics[scale=0.33]{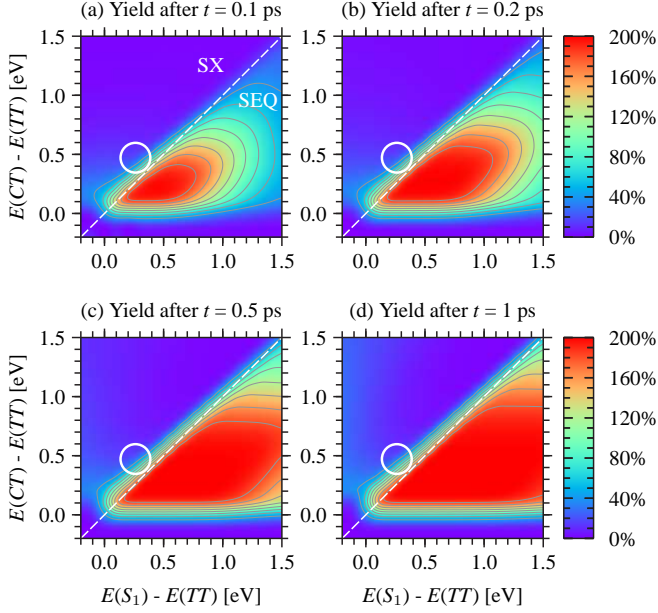}
\caption{
The same as in Fig.~\ref{fig:2dyield} but for population dynamics calculated by the NIBA-type master equation,
which is perturbative to second order in the electronic couplings, $V_{ij}$.
}
\label{fig:2dyield_niba}
\end{figure}

In light of this analysis, one should expect that an alternative master equation that is perturbative to
second-order in the electronic coupling (akin to F{\" o}rster theory or the noninteracting blip approximation (NIBA),
as in the work of Teichen and Eaves in Ref.~\onlinecite{tei12})
would only predict sequential, and not superexchange, CT-mediated
fission mechanisms.  In Fig.~\ref{fig:2dyield_niba}, we show the fission dynamics predicted by the NIBA-like master
equation,
\begin{equation}
\frac{dP_i(t)}{dt} = \sum_j \int_0^t ds K_{ij}(t,s) P_j(s),
\end{equation}
where
\begin{equation}
K_{ij}(t,s) = \frac{2 |V_{ij}|^2}{\hbar^2} \textrm{Re} \left\langle \exp\left({-i\hat{H}^{tot}_{ii}t/\hbar}\right) \exp\left({i\hat{H}^{tot}_{jj}s/\hbar}\right) \right\rangle_{ph},
\end{equation}
and $\hat{H}^{tot}_{ii} = \langle i | \hat{H}^{tot} | i\rangle$,
clearly demonstrating that such master equations
incorrectly predict no fission if $E(CT) > E(S_1)$.

\begin{figure}[t]
\centering
\includegraphics[scale=1.0]{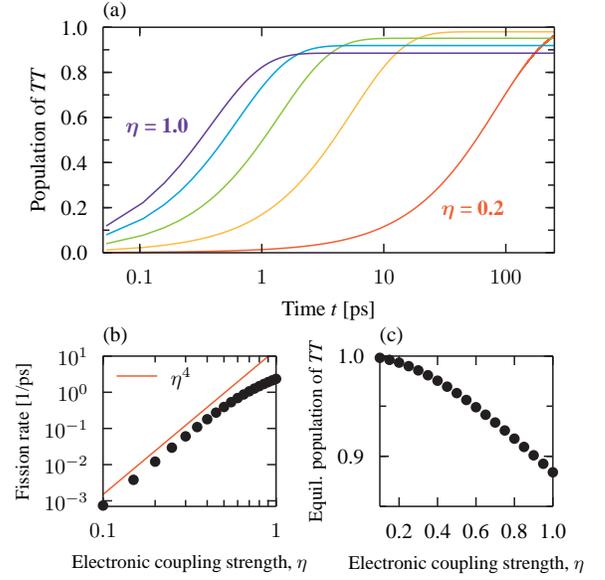}
\caption{
Dramatic slowing down of singlet fission dynamics for decreasing electronic coupling strength $\eta$ (a); note that the time axis
is in log-scale.  The numerically extracted fission rate obeys the predicted superexchange scaling $k \sim \eta^4$ (b), however
the equilibrium population of $TT$ decreases with increasing coupling, due to enhanced mixing with non-$TT$ states (c).
}
\label{fig:scaling}
\end{figure}

To check the validity of the derived fourth-order scaling, we introduce a dimensionless parameter, $\eta$, which characterizes the electronic
coupling strength.  Specifically, we replace $V_{ij} \rightarrow \eta V_{ij}$, and consider the limit $\eta \rightarrow 0$,
for which a superexchange mechanism predicts $k \propto \eta^4$.
In Fig.~\ref{fig:scaling}(a), we see that the rate of fission decreases drastically as $\eta \rightarrow 0$, approaching
a 100 ps timescale for $\eta = 0.2$.  The fission rates (obtained by a numerical fit of the $TT$ population growth) are then
plotted in log-log scale, Fig.~\ref{fig:scaling}(b), very clearly confirming the superexchange scaling $k \propto \eta^4$.
Another interesting feature is apparent in the long time dynamics of Fig.~\ref{fig:scaling}(a). To quantify this behavior,
in Fig.~\ref{fig:scaling}(c), we plot the equilibrium population of $TT$,
obtained as $P_{TT}(t\rightarrow \infty)$ (equivalently $Z_{el}^{-1} \langle TT | \exp(-\hat{H}_{el}/k_BT) | TT \rangle$)
as a function of the electronic coupling strength $\eta$.  Clearly, for increasing $\eta$, the equilibrium population shows
a noticeable decline, which is straightforwardly explained: for stronger values of coupling, the zeroth-order diabatic states
are more strongly mixed, such that the lowest energy adiabatic eigenstate develops a larger fraction of non-$TT$ states, effectively
depleting the diabatic $TT$ population.

The mechanistic features evinced in this and the previous section constitute the main results of this paper.  Namely,
a superexchange two-electron-transfer phenomenon, utilizing {\em virtual} states in the $CT$ vibronic manifold, is entirely consistent with
the observed features of singlet fission in pentacene.  As such, a CT-mediated singlet fission mechanism cannot be ruled out based solely
on the argument that CT states are too high in energy.  From a more technical point of view, only a dynamical master equation which can account
for fourth-order effects in the electronic couplings, such as Redfield theory, is able to correctly predict this behavior.  Before considering
the feasibility of a direct mechanism, mediated by a true two-electron coupling matrix element, we investigate in the next section the role played by the
phonon bath degrees of freedom.

\subsection{The effect of the bath}\label{ssec:bath}

Within the weak system-bath coupling approximation of Redfield theory, the adiabatic population transfer rates
$k_{\alpha\rightarrow \beta} \equiv R_{\beta\beta\alpha\alpha}$
are given by
\begin{subequations}\label{eq:rates}
\begin{align}
k_{\alpha\rightarrow\beta} &= \hbar^{-1} C_{\alpha\beta} J(\omega_{\alpha\beta}) n(\omega_{\alpha\beta}) \\
k_{\beta\rightarrow\alpha} &= \hbar^{-1} C_{\alpha\beta} J(\omega_{\alpha\beta}) \left[ n(\omega_{\alpha\beta}) + 1\right] 
\end{align}
\end{subequations}
where
$C_{\alpha\beta} = \sum_i \left| \langle \beta | i\rangle \right|^2 \left|\langle \alpha | i \rangle \right|^2$
arises from the change of basis,
$n(\omega) = \left[\exp(\hbar\omega/k_B T)-1\right]^{-1}$ is the Bose-Einstein distribution,
and we assume $\omega_{\alpha\beta} = \omega_\alpha-\omega_\beta > 0$. Clearly these transfer rates satisfy
the detailed balance condition,
\begin{equation}
\frac{k_{\alpha\rightarrow\beta}}{k_{\beta\rightarrow\alpha}} = \exp\left(\hbar\omega_{\alpha\beta}/k_B T\right).
\end{equation}
Furthermore, the transfer rates are directly proportional to the spectral density evaluated at the eigenvalue
energy difference.  Physically, the transfer rate between two states depends on the availability of strongly-coupled
phonon modes at the required energy difference as well as their thermal occupancy, so that phonon absorption
and emission facilitates the electronic energy transfer.

\begin{figure}[t]
\centering
\includegraphics[scale=1.0]{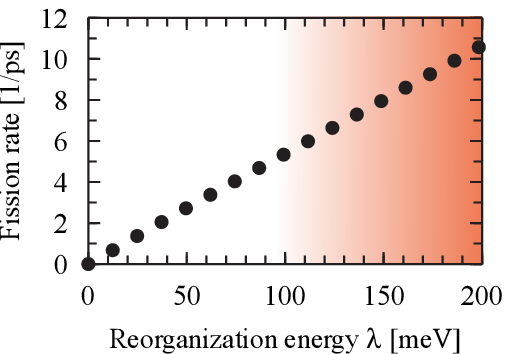}
\caption{
Calculated fission rate for a pentacene dimer with varying system-bath coupling, quantified by the reorganization
energy, $\lambda$.  Secular, Markovian Redfield theory (filled circles) predicts a linear dependence, which
is known to be accurate for small $\lambda$ but becoming more inaccurate for large $\lambda$ (indicated by the shaded
region).  Realistic values for pentacene are $\lambda \approx$ 50--150 meV, which reliably predicts a fission
rate $k \approx$ 2--10 ps$^{-1}$ i.e. $\tau \approx$ 100--500 fs, in reasonably good agreement with experimental
rates of 80--200 fs.
}
\label{fig:lambda}
\end{figure}

The fission rate will clearly depend on the parameters and functional form of the spectral density, $J(\omega)$.
As discussed previously, the chosen reorganization energy $\lambda = 50$ meV is conservatively small, and so here
we analyze the dependence of the fission rate on the strength of the system-bath coupling, quantified by $\lambda$.
Because $J(\omega) \propto \lambda$ for any form of the spectral density, Redfield theory predicts the trivial
linear dependence $k_{\rm fiss} \propto \lambda$ via Eqs.~(\ref{eq:rates}); see Fig.~\ref{fig:lambda}.  While this is indisputably
the correct behavior in the small $\lambda$ limit, it becomes incorrect for $\lambda$ sufficiently large\cite{ish09_jcp1}.
This breakdown is visually suggested in Fig.~\ref{fig:lambda}, with a shaded region indicating at what point the 
theory may become inaccurate, initially quantitatively but ultimately qualitatively\footnote{The reorganization energy
at which Redfield theory breaks down was estimated by comparison to numerically exact results on small
model systems, presented in our previous paper.  Analogous exact calculations on the identical five-level system
considered here are difficult to converge completely due to the high-frequency bath, but also confirm that Redfield
theory breaks down near $\lambda = 100$ meV.}

\begin{figure}[b]
\centering
\includegraphics[scale=1.0]{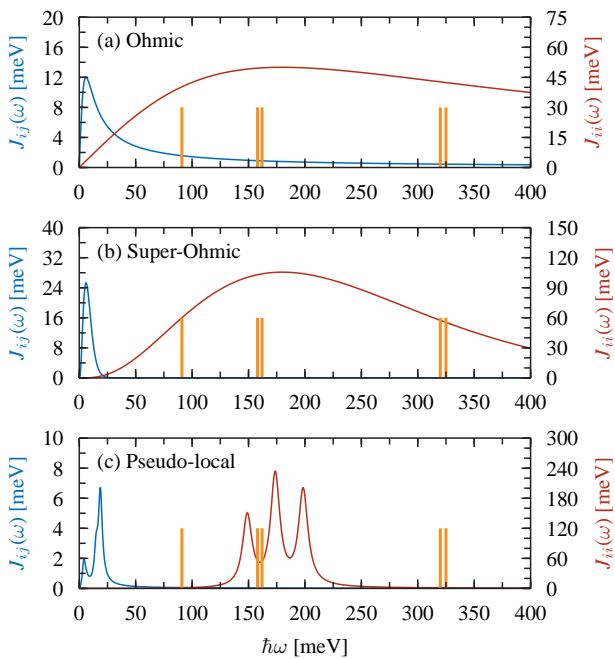}
\caption{
Three different forms of the spectral density investigated here, along with the electronic eigenvalue differences for pentacene (orange vertical sticks).
The overlap between these energy differences and the spectral density, i.e. the ability to absorb and emit resonant phonons,
largely determines the rate of population transfer and hence singlet fission.
}
\label{fig:spec}
\end{figure}

We conclude this section by considering different functional forms of the spectral density, although we stress that
the Redfield formalism is equally accurate and efficient for any form of the spectral density, including ones determined
numerically.  In addition to the Ohmic form utilized so far, we will also consider
a super-Ohmic spectral density of Debye-type phonons with exponential cutoff,
\begin{equation}
J^{SO}_{ij}(\omega) = \frac{27 \pi \lambda_{ij}}{2\Omega_{ij}^3} \omega^3 \exp(-3 \omega/\Omega_{ij}),
\end{equation}
with $\lambda = 50$ meV and $\Omega = 180$ meV as before.
Lastly, we will consider a broadened stick-spectrum of pseudo-local phonon modes\cite{hsu85},
\begin{equation}
J^{PL}_{ij}(\omega) = \frac{1}{\pi} \sum_k \frac{\lambda_{k,ij} \Gamma_{k,ij} \omega }{(\omega-\Omega_{k,ij})^2 + \Gamma_{k,ij}^2},
\end{equation}
with total reorganization energy $\lambda_{ij} = \sum_k \lambda_{k,ij}$.
The latter has three phonon modes, chosen based on the results presented by Girlando et al.\cite{gir11},
with $\lambda_k = \{ 15, 20, 15 \}$ meV ($\sum_k \lambda_k = 50$ meV) at frequencies $\hbar \omega_k = \{ 150, 175, 200 \}$ meV, with uniform broadening
$\hbar\Gamma_k = 50$ meV.

As just discussed, the rate is largely determined by the overlap of the spectral density with the
eigenvalue energy differences.  In Fig.~\ref{fig:spec}, we plot these three different spectral densities along with the
energy differences, visually portraying which modes mediate which electronic transitions.  Clearly, the overlap is most
uniform for the Ohmic and least uniform for the pseudo-local spectral density: while the pseudo-local spectrum greatly
enhances some electronic transition rates, it greatly diminishes other, in particular those with relatively small
energy differences.  In Fig.~\ref{fig:spec} we also show a possible Peierls-type (off-diagonal) spectral density based
on the crystal phase calculations of Girlando et al.\cite{gir11}, though
we reiterate that such coupling was not employed in the results presented here.  Based on the overlap argument given above,
it is visually apparent why these low frequency fluctuations do not efficiently mediate the fission process, as we have
found in dynamics calculations not shown.

\begin{figure}[t]
\centering
\includegraphics[scale=1.0]{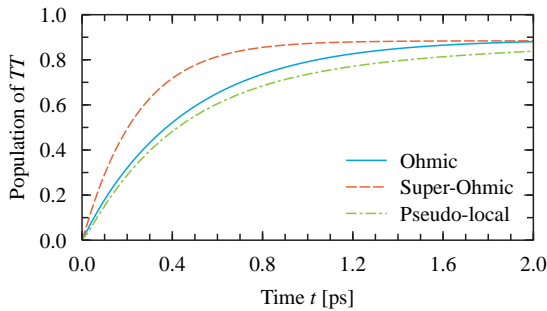}
\caption{
Singlet fission population dynamics for the three spectral densities depicted in Fig.~\ref{fig:spec}.  
}
\label{fig:baths_dyn}
\end{figure}

The fission dynamics due to these various spectral densities are found to be only mildly different, as shown
in Fig.~\ref{fig:baths_dyn}.  This behavior is due largely to the uniformly strong coupling to phonon modes near 160 meV
as these phonon modes mediate the important $S_1 \Rightarrow (S_1 \leftrightarrow CT)$ and $(S_1 \leftrightarrow CT)\Rightarrow TT$
transitions, see Fig.~\ref{fig:diabatic}.  Based on these results, one can in principle imagine devising clever ways to engineer either
the electronic spectrum or the phonon spectrum to realize maximal singlet fission rates.  In practice, an accurate
determination of the ``correct'' spectral density is very difficult, but because it can clearly shape the observed
electronic dynamics, we consider this an important topic worthy of further study.  

\subsection{Direct mechanism}

While the results so far have focused on the features of a CT-mediated singlet fission mechanism, we now briefly
investigate the feasibility of a direct-coupling pathway.  Ultimately, we find that the required two-electron
integral which couples $S_1$ to $TT$ is too small to explain efficient singlet fission in pentacene dimers
Of course, it remains to repeat the analysis in crystals, for which
the multiexcitonic density of states is enhanced as compared to the dimer.
This calculation will be reported in a future paper.

Using the methodology described in Sec.~\ref{sec:method}, we calculated the direct coupling matrix element,
\begin{equation}
\langle S_1 S_0 | \hat{H}_{el} | T T \rangle = \sqrt{\frac{3}{2}}\big\{ (L_A L_B | H_B L_A) - (H_A H_B | L_B H_A) \big\},
\end{equation}
where
\begin{equation}
(ij|kl) = \int d^3\vr_1 \int d^3\vr_2 \phi^*_i(\vr_1) \phi^*_j(\vr_2) r_{12}^{-1} \phi_k(\vr_1) \phi_l(\vr_2),
\end{equation}
finding it to be \textit{less than 1 meV}.  This value (which is in agreement with more accurate calculations
using block-localized DFT\cite{proxx,chaxx}) should be contrasted with the \textit{effective}
superexchange coupling, Eq.~(\ref{eq:supercoupling}), which is approximately 50 meV for the pentacene dimer.

\begin{figure}[t]
\centering
\includegraphics[scale=1.0]{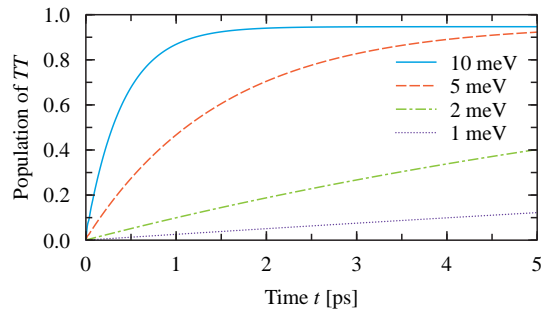}
\caption{
Singlet fission population dynamics in the absence of $CT$ states, for varying values of the direct electronic
coupling element given in the legend.  Sub-picosecond fission is only observed for the unphysically large value
of 10 meV, to be contrasted with theoretical estimates ranging from 5 to less than 1 meV.
}
\label{fig:direct}
\end{figure}

These qualitative arguments are numerically confirmed in Fig.~\ref{fig:direct}, where we show the singlet fission
dynamics in the absence of a CT-mediated pathway, i.e. the electronic coupling to all CT states is zero, thereby
reducing the number of relevant states to three.
In particular, we use $E(S_1)-E(TT) = 250$ meV, $\lambda = 50$ meV, and bath frequency and temperature as before.  
By varying the strength of the two-electron coupling, we observe that the direct mechanism only becomes competitive
at unrealistically large values.  As such, we conclude that, while feasible, the direct
singlet fission mechanism is subdominant as compared to CT-mediated superexchange, at least for pentacene dimers.

\subsection{Covalently linked dimer}

\begin{figure}[b]
\centering
\includegraphics[scale=0.8]{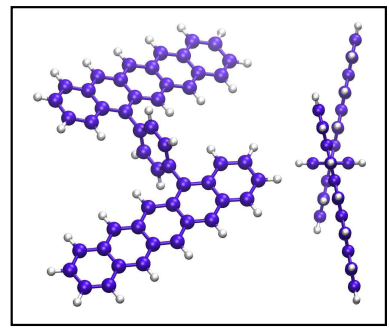}
\caption{
Two different views of a covalently linked pentacene dimer akin to the tetracene dimers of Refs.~\onlinecite{mul06,mul07}.
The electronic couplings are estimated to be significantly smaller than
in the crystal phase dimer pairs investigated above, predicting a much smaller rate of fission.
}
\label{fig:bardeen}
\end{figure}

In this final section, we briefly investigate a molecular geometry that is different from those considered before.  In particular,
inspired by the synthesized tetracene molecules of M{\" u}ller et al.\cite{mul06,mul07}, we consider the analogous
pentacene dimer shown in Fig.~\ref{fig:bardeen}.
In Refs.~\onlinecite{mul06,mul07}, the authors found that less than $1\%$ of singlets
underwent fission, which is in stark contrast to the relatively high fission yield of crystalline tetracene.

The covalently bound dimer pair contains a benzene linker, which
can facilitate a through-bond coupling in addition to the through-space coupling available in the crystal.
We can straightforwardly estimate the through-space coupling with the methodology described above by excising the benzene linker and
terminating the dangling bonds with hydrogens. This calculation yields one-electron coupling matrix elements, $t_{ij}$,
of approximately 0.02 meV, which is three orders of magnitude smaller than in the native crystal conformation.  The mixed through-bond
couplings, $t_{HL}$ and $t_{LH}$, are not so easily obtained, but the HOMO-HOMO and LUMO-LUMO couplings can be estimated using
the energy-splitting method alluded to previously.  For this calculation, the benzene linker is retained in the molecule, and the splitting
of the HOMO and LUMO orbitals is assigned to twice the respective coupling.  We calculate this through-bond coupling to be on the order
of 20 meV, significantly larger than the through-space contribution, but still 4-10 times smaller than the through-space coupling
of the crystalline dimer pairs investigated above.  In light of the
scaling analysis presented in Sec.~\ref{ssec:sxscaling}, it is clear that such a reduction will yield a significantly slower rate of
fission, increasing the timescale from sub-picosecond to as much as one nanosecond, extending the timescale over which fission
must compete with other decay mechanisms.
The torsional modes about the benzene linker, which couple to these off-diagonal electronic hopping matrix elements, 
are expected to be of relatively low frequency and thus would not qualitatively affect the fission dynamics, similar to the analogous observation
we made for the crystal.
This example calculation
provides a possible explanation for the low observed fission yield\cite{mul06,mul07} and encourages efforts to explore electronic coupling
effects through a combination of covalent bonding and optimization of geometric orientation.

\section{Conclusions}\label{sec:conc}

In this and the previous work, we have presented and applied a unified microscopic theoretical framework for the investigation
of singlet fission electronic structure and dynamics.  We have emphasized the role played by molecular vibrations
or phonons in mediating non-resonant excited state energy transfer.  Such finite temperature relaxation mechanisms
are responsible for population transfer, coherence dephasing, and eventual thermalization.  These processes
may be numerically studied and quantified using a perturbative quantum master equation subject to its regime
of validity.  We have justified and pursued Redfield theory for singlet fission chromophore systems,
advocating its favorable trade-off between accuracy and efficiency.
Within our framework, we have performed a thorough investigation of singlet fission in pentacene dimer systems,
in particular investigating the real-time fission dynamics.  In principle, all parameters needed to carry out the real-time
quantum dynamics at the level detailed in this work may be estimated from microscopic considerations.
Instead, in this work we have carried out a systematic variation of several parameters which influence singlet fission
efficiency, providing insight into the underlying mechanistic details which are beyond the reach of experiment.

Most notably, we have provided evidence for a CT-mediated superexchange mechanism in pentacene dimers, which is more efficient
than a direct mechanism, even in the presence of very high-energy intermediate CT states.  These results
are in stark contrast with previous theoretical predictions\cite{zim11,tei12}.  However, our superexchange mechanism
bears many similarities to the conventional direct mechanism and it is in some sense a hybrid between the two
competing mechanisms.  We have investigated the way fission rates and yields are modified by shifting the electronic energy levels 
and scaling the electronic couplings.  Our study of the role of the phonon bath properties underscores
the importance of having resonant phonon frequencies to mediate efficient energy transfer.
Lastly, we have shown that, at least in pentacene dimers, the direct coupling singlet fission
pathway is extremely inefficient and would require an unphysically large two-electron matrix element to compete with
the mediated superexchange mechanism whose timescale is already in very good agreement with experimentally observed
fission rates in pentacene crystals.

It should be emphasized, however, that it may be dangerous to draw conclusions about singlet fission in the bulk based
on the calculations for idealized dimers presented here. While our work clearly shows that the existence of CT
states that are energetically high-lying in no way obviates their importance in singlet fission, it does not prove
that superexchange is relevant for bulk pentacene or other singlet fission materials. Further, the possibility
exists that an enhancement of the direct mechanism may take place in bulk materials due to the augmented density
of multi-exciton states expected in large clusters and bulk solids and films. Our preliminary calculations on 
larger pentacene clusters based on the formalism presented here, which allows for the efficient simulation of hundreds of
quantum states, suggest that many of our conclusions are unaltered, in particular that CT states are intrinsically connected to fission and 
the predominant mechanism is not a direct two-electron process.  This work will be presented
in a forthcoming paper.

\begin{acknowledgments}
We thank Eran Rabani for the use of his two-electron integral FFT code.
This work was supported in part by the Center for Re-Defining Photovoltaic Efficiency through Molecule Scale
Control, an Energy Frontier Research Center funded by the US Department of Energy, Office of Science, Office of
Basic Energy Sciences under Award Number DE-SC0001085.
This work was carried out in part at the Center for Functional Nanomaterials, Brookhaven National Laboratory,
which is supported by the U.S. Department of Energy, Office of Basic Energy Sciences under Contract No. DE-AC02-98CH10886 (M.S.H).
T.C.B. was supported in part by the Department of Energy Office of Science Graduate Fellowship Program (DOE SCGF),
made possible in part by  the American Recovery and Reinvestment Act of 2009, administered by ORISE-ORAU
under Contract No. DE-AC05-06OR23100.
\end{acknowledgments}

\bibliography{fission}

\end{document}